\documentclass[preprintnumbers,nofootinbib,noshowpacs,eqsecnum,prd,superscriptaddress,letterpaper]{revtex4}

\usepackage{graphicx}
\usepackage{amsmath,amssymb}
\usepackage{url}
\usepackage{multirow}
\usepackage{hyperref,color}
\graphicspath{{figs/}}
\usepackage{xcolor}

\begin{document}

\title{Indirect and monojet constraints on scalar leptoquarks}
\author{Alexandre Alves}
\email{aalves.unifesp@gmail.com}
\affiliation{Departamento de F\'isica, Universidade Federal de S\~ao Paulo, UNIFESP,
Diadema, S\~ao Paulo, Brazil}
\author{Oscar J. P. \'Eboli}
\email{eboli@if.usp.br}
\affiliation{Instituto de F\'isica, Universidade de S\~ao Paulo,
 S\~ao Paulo, Brazil}
\author{Giovanni Grilli di Cortona}
\email{ggrillidc@fuw.edu.pl}
\affiliation{Instituto de F\'isica, Universidade de S\~ao Paulo,
 S\~ao Paulo, Brazil}
\affiliation{Institute of Theoretical Physics, Faculty of Physics, University of Warsaw, ul.~Pasteura 5, PL--02--093 Warsaw, Poland}
\author{Roberto R. Moreira}
\email{robertor@fma.if.usp.br}
\affiliation{Instituto de F\'isica, Universidade de S\~ao Paulo,
 S\~ao Paulo, Brazil}
%

\begin{abstract}
  
  We obtain constraints on first- and second-generation scalar
  leptoquarks using the available data on dilepton (Drell-Yan) and
  monojet searches at the CERN Large Hadron Collider. Assuming that
  the leptoquark interactions respect the Standard Model gauge
  symmetries as well as lepton and baryon numbers, we show that the
  study of dilepton production enlarges the exclusion region on the
  mass and coupling plane with respect to the pair production searches
  for first-generation leptoquarks. Moreover, the monojet channel
  leads to a larger excluded parameter region for moderate to large
  values of the leptoquark Yukawa coupling than the presently
    available experimental results.

\end{abstract}

\maketitle

\section{Introduction}
\label{sec:intro}

The CERN Large Hadron Collider (LHC) has gathered a large amount of
data that has been used not only to probe the Standard Model (SM), but
also to look for new physics. The SM does not have any state that
couples to lepton-quark pairs, therefore, the discovery of such a
state would be an indubitable sign of Physics beyond the SM. In fact this is
an ongoing search by the LHC collaborations; see, for instance,
Refs.~\cite{Aaboud:2016qeg, Sirunyan:2018btu}. \smallskip

Leptoquarks are particles which interact with quark-lepton pairs.
They appear in a plethora of models where quarks and leptons are
treated in the same footing. The first appearance of leptoquarks was
in the Pati-Salam model~\cite{Pati:1973uk,Pati:1974yy} and they are
also present in Grand Unified Theories~\cite{Georgi:1974sy}, composite
models~\cite{Schrempp:1984nj}, or in supersymmetric models with
violation of $R$ parity~\cite{Barbier:2004ez}.
Leptoquarks modify the low-energy physics which, in turn, leads to
constraints on their masses and couplings. For instance, leptoquarks
might contribute to the decay of mesons~\cite{Shanker:1981mj}, induce
flavor changing neutral currents (FCNC)~\cite{ Pati:1974yy,
  Shanker:1981mj}, give rise to lepton flavor violating
decays~\cite{Dorsner:2016wpm}, and contribute to the radiative
correction to $Z$ physics~\cite{Mizukoshi:1994zy,
  Bhattacharyya:1994ig}.
Recently some anomalies were observed that point towards lepton flavor
universality violation in neutral- and charged-current
processes~\cite{Lees:2012xj,Aaij:2015yra, Huschle:2015rga,
  Sato:2016svk, Amhis:2016xyh,
  Aaij:2013qta,Aaij:2015oid,Ciuchini:2015qxb, Aaij:2014ora,
  Aaij:2017vbb}. One possible explanation to the departures from the
SM is the existence of leptoquarks with masses ${\cal O}$(TeV); see,
for instance, Refs.~\cite{Dorsner:2013tla,
  Sakaki:2013bfa,Calibbi:2015kma, Bauer:2015knc, Barbieri:2015yvd,
  Becirevic:2016oho,Crivellin:2017zlb, Assad:2017iib, Calibbi:2017qbu,
  Blanke:2018sro, Crivellin:2018yvo, Biswas:2018snp, Mandal:2018kau,
  Biswas:2018iak, Aebischer:2018acj, Fornal:2018dqn} and references
therein.  \smallskip

Since leptoquarks carry color, they can be pair produced via
QCD~\cite{Hewett:1987yg} at the LHC and the main advantage of this
channel is that the production cross section depends only on the
leptoquark mass.  Presently most of the LHC searches for leptoquarks
are based on their pair production.  The LHC RUN 1 data excludes
first- (second-) generation scalar leptoquarks with masses smaller
than 1050 (1080) GeV ~\cite{Aad:2015caa, Khachatryan:2015vaa} provided
the leptoquark decays exclusive into a charged lepton and a jet. The
already available results from LHC RUN 2 expands these limits to 1435
GeV and 1520 GeV for first- and second-generation leptoquarks,
respectively~\cite{Sirunyan:2018btu, Sirunyan:2018ryt}.\smallskip

Leptoquarks can also be produced singly in association with a lepton
through its coupling to a quark-lepton
pair~\cite{Eboli:1987vb,Mandal:2015vfa,Bansal:2018eha,Schmaltz:2018nls}. In fact, this process has a
larger phase space than the leptoquark pair production and for
${\cal O}(1)$ couplings and 100\% branching ratio into charged lepton
plus up quark, the bounds (${\cal O} (1.3$--1.7) TeV) from the Run I
data of the LHC can be even stronger than the pair production
ones~\cite{Khachatryan:2015qda}.
At the LHC it is also possible to look for indirect signs of
leptoquarks in the production of charged lepton
pairs~\cite{Eboli:1987vb, Raj:2016aky,
  Schmaltz:2018nls,Bansal:2018dge}.  \smallskip

Leptoquarks coupling to a neutrino and a quark 
give rise to monojet events due to their single production.  In this
case, the leptoquark decay leads to a hard missing transverse energy
spectrum peaking around half the leptoquark mass.  
It is interesting to notice that models where leptoquarks couple to
dark matter~\cite{Allanach:2015ria} with sizable branching ratios
could also be constrained by these monojet searches. At the same time
they alleviate other constraints due to the diminished leptoquark
branching ratio to SM lepton plus quark.
Moreover, the LHC collaborations also searched for leptoquarks in the
mono-bottom ~\cite{Lin:2013sca,CMS:2016uxr,Aad:2014vea} and mono-top
quark~\cite{Sirunyan:2018gka,Sirunyan:2018dub} channels which have
been explored in the search for dark matter. These channels constitute
alternatives to pair production in the case where the third-generation
leptoquarks are too heavy. \smallskip

If leptoquarks are too heavy to be pair produced at the LHC, single
production and indirect effects in lepton pair production remain as
alternatives to search for these particles~\cite{Eboli:1987vb}.  In
this work we study the attainable bounds on first- and
second-generation scalar leptoquarks using the available data on
lepton pair production and the monojet searches for dark
matter~\cite{CMS:2017tbk}. Our results depend on the leptoquark mass
as well as its Yukawa coupling to quark-lepton pairs. As a consequence, they
will yield more information on the leptoquark properties if a
signal is observed, complementing the leptoquark pair production searches.
We show that the study of dilepton production enlarges the exclusion
region in the plane mass and coupling with respect to the pair
production searches.
Additionally, the monojet channel for first-generation leptoquarks
also leads to a larger exclusion region than the canonical
$\ell^+ \ell^- jj$ and $\ell^\pm \nu jj$ (with $\ell =e$ and $\mu$)
topologies for moderate and large values of the leptoquark Yukawa
coupling.
The monojet channel is also able to probe a larger fraction of the
parameter space than the $jj E^{\rm miss}_T$ final
state~\cite{Sirunyan:2018kzh}.
\smallskip

\section{Analyses Framework}
\label{sec:frame}

In order to describe the leptoquark interactions at the presently
available energies we assume that leptoquarks are the only accessible
states from an extension of the SM. Moreover, we impose that their
interactions respect the SM gauge symmetry
$SU(3)_c \otimes SU(2)_L \otimes U(1)_Y$. Furthermore, due to the
stringent bounds coming from proton lifetime experiments the
leptoquark interactions are required to respect baryon and lepton
numbers. Within these hypotheses, there are five distinct
possibilities for scalar leptoquarks~\cite{Buchmuller:1986zs}: two
$SU(2)_L$ singlet states ($S_1$ and $\tilde{S}_1$), two doublet states
($R_2$ and $\tilde{R}_2$) and one triplet ($S_3$), whose interactions
with quarks and leptons are
\begin{eqnarray}
  {\cal L} _{\rm eff}  
=&&~ \left ( g_{\text{1L}}~ \bar{q}^c_L~ i \tau_2~ \ell_L +
g_{\text{1R}}~ \bar{u}^c_R~ e_R \right )~ S_1
+ \tilde{g}_{\text{1R}}~ \bar{d}^c_R ~ e_R ~ \tilde{S}_1
+ g_{3L}~ \bar{q}^c_L~ i \tau_2~\vec{\tau}~ \ell_L \cdot \vec{S}_3
\; ,
\nonumber 
\\
 &&+  ~ h_{\text{2L}}~ R_2^T~ \bar{u}_R~ i \tau_2 ~ \ell_L
+ h_{\text{2R}}~ \bar{q}_L  ~ e_R ~  R_2
+ \tilde{h}_{\text{2L}}~ \tilde{R}^T_2~ \bar{d}_R~ i \tau_2~ \ell_L
\; ,
\label{eq:leff} 
\end{eqnarray}
where $q_L$ ($\ell_L$) stands for the left-handed quark (lepton) doublet,
$u_R$, $d_R$, and $e_R$ are the singlet components of the fermions,
$\tau_i$ are the Pauli matrices, and we denote the charge conjugated
fermion fields by $\psi^c=C\bar\psi^T$.  Moreover, we omit the flavor
indices of the fermions. The couplings of the leptoquarks to the
electroweak gauge bosons is determined by the
$SU(3)_c \otimes SU(2)_L \otimes U(1)_Y$ gauge invariance.  \smallskip

At low energies, leptoquarks give rise to flavor changing neutral
current (FCNC) processes~\cite{ Pati:1974yy, Shanker:1981mj,
  Buchmuller:1986iq} that are very constrained. In order to avoid
stringent bounds, we assume that the scalar leptoquarks couple only to
a single generation of quarks and just one of leptons. In fact, we
consider that the leptoquarks interact with the same generation of
quarks and leptons. Notwithstanding, there is a residual amount of
FCNC due to the mixing in the quark sector that leads to bounds on
leptoquark couplings to the first two
generations~\cite{Davidson:1993qk,Carpentier:2010ue}. 
Moreover, data on decays of pseudo-scalar mesons, like the pions, put
stringent bounds on leptoquarks unless their couplings are chiral –-
that is, they are either left- or
right-handed~\cite{Shanker:1981mj,Davidson:1993qk, Carpentier:2010ue}.
As a rule of a thumb, the low-energy data constrain
first-generation-leptoquark masses to be larger than 0.5--1.0 TeV for
leptoquark Yukawa couplings of the order of the proton charge
$e$~\cite{Leurer:1993em}. \smallskip

In our analysis  we focus on leptoquarks that  conserve fermion number
$F=3B+L$, {\em i.e.} the states $R_2$ and $\tilde{R}_2$.  For the sake
of  simplicity,  we  assume  that leptoquarks  belonging  to  a  given
$SU(2)_L$ multiplet are degenerate.   We study the following scenarios
that satisfy the described low-energy bounds:
\begin{enumerate}

\item $\tilde{R}_2$ coupling only to the first generation, {\em i.e.}
  only to down quarks, electrons and its neutrino. We label this
  scenario as $\tilde{R}_2^d$.

\item $R_2$ coupled to the first generation and exhibiting only
  left-handed interactions ($h_{2R}=0$). In this case the leptoquarks
  interact with up quarks, electrons and the respective neutrino. We
  refer to this scenario as $R_2^{L}$.

\item $\tilde{R}_2$ coupling only to the second generation -- that is
  to strange quarks, muons and the corresponding neutrinos. We call
  this scenario $\tilde{R}_2^s$.

\end{enumerate}
These scenarios allows us to obtain bounds on other leptoquarks. For
instance, the constraints on $R_2$ (second scenario) originating from
the dilepton data can be easily translated into bounds on the singlet
$S_1$ or the third component of the triplet $\vec{S}_3$. On the other
hand, the dileptons limits in the first scenario can be readily
transposed to $\tilde{S}_1$ and one component of
$\vec{S}_3$. Furthermore, the $\tilde{R}_2^s$ case is the one leading
to the strongest constraints on second-generation leptoquarks.
\smallskip


In order to study leptoquarks at the LHC we consider their indirect
effects in Drell-Yan processes~\cite{Eboli:1987vb}
\begin{equation}
p p \to e^+ e^-/ \mu^+ \mu^- + X
\label{eq:dy}
\end{equation}
as well as in monojet searches
\begin{equation}
p p \to j + E_T^{\rm miss}  \;\;.
\label{eq:monoj}
\end{equation}

In the Drell-Yan 8 TeV analyses we use the data in
Ref.~\cite{Aad:2016zzw} that contain the $e^+e^-$ and $\mu^+\mu^-$
invariant mass distribution in the fiducial region defined by the
leading (subleading) lepton having $p_T^\ell > 40$ GeV (30 GeV) and
both leptons within the rapidity range $|\eta| < 2.5$. These results
are given at QED Born level which simplifies the comparison with the
leptoquark predictions.
Since the 13 TeV Drell-Yan analyses have not been released by ATLAS
and CMS we use the data on searches for new resonances decaying into
lepton pairs given in Refs.~\cite{Aaboud:2017buh, Sirunyan:2018exx}.
More specifically, the data used was\footnote{We merged the two last
  bins in Ref.~\cite{Aaboud:2017buh} to ensure gaussianity. The last
  bin of Ref.~\cite{Sirunyan:2018exx} (denoted by $1800_+$) is for
  $m_{ee}(m_{\mu\mu})> 1800$ GeV.}
\begin{center}
\begin{tabular}{ l|lclcc}
\hline 
Channel  & kinematical range & \# bins  & Int. Lum. & & Data set
\\ [0mm]
\hline
$e^+ e^-$ & $ 116 < m_{ee} < 1500$ GeV & 12 & 20.3 fb$^{-1}$
& & ATLAS 8 TeV~\cite{Aad:2016zzw}  Table 6 
\\[0mm]
$\mu^+ \mu^-$ & $ 116 < m_{\mu\mu} < 1500$ GeV & 12 & 20.3 fb$^{-1}$
& & ATLAS 8 TeV~\cite{Aad:2016zzw}  Table 9 
\\[0mm]
\hline
$e^+ e^-$ & $ 80 < m_{ee} < 6000$ GeV & 9 & 36.1 fb$^{-1}$
& & ATLAS 13 TeV~\cite{Aaboud:2017buh}  Table 6 
\\[0mm]
$\mu^+ \mu^-$ & $ 80 < m_{\mu\mu} < 6000$ GeV & 9 & 36.1 fb$^{-1}$
& & ATLAS 13 TeV~\cite{Aaboud:2017buh}  Table 7
\\[0mm]
\hline
$e^+ e^-$ & $ 120 < m_{ee} < 1800_+$ GeV & 6 & 36. fb$^{-1}$
& & CMS 13 TeV~\cite{ Sirunyan:2018exx}  Table 2
\\[0mm]
$\mu^+ \mu^-$ & $ 120 < m_{\mu\mu} < 1800_+$ GeV & 6 & 36. fb$^{-1}$
& & CMS 13 TeV~\cite{ Sirunyan:2018exx}  Table 3
\\[0mm]
\hline

\end{tabular}
\end{center}

As for the monojet analysis, we use the CMS results~\cite{CMS:2017tbk}
at 13 TeV with an integrated luminosity of 35.9 fb$^{-1}$.  The CMS
analysis was done as a counting experiment in 22 independent signal
regions satisfying $E_T^{\rm miss} > 250$ GeV, $p_T^{\rm jet} > 100$
GeV and $|\eta_j| < 4.5$.  For each of the above experiments and
channels, we extract from the experimental publications the observed
event rates in each bin $N^{i}_{\rm obs}$, the background expectations
$N^{i}_{\rm bkg}$, and the SM predictions $N^{i}_{\rm SM}$, as well as
the statistical and systematic errors. \smallskip

\begin{figure}[t!]
\centering
\includegraphics[width=0.225\textwidth]{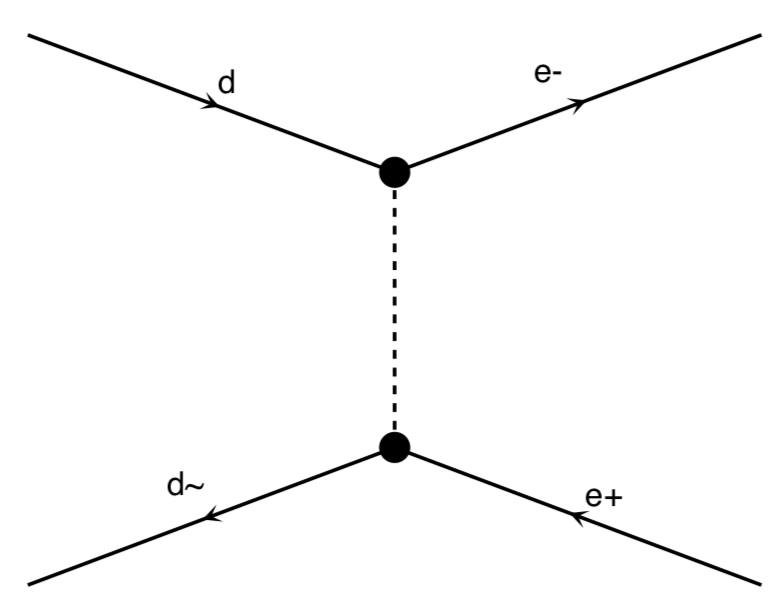}
~\includegraphics[width=0.45\textwidth]{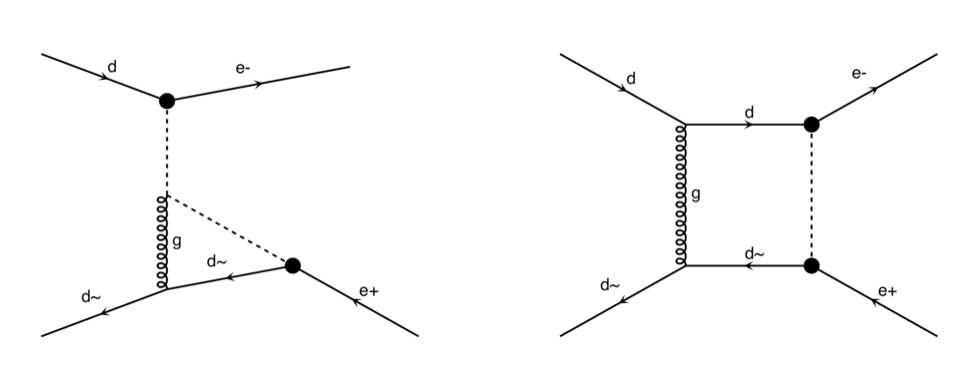}
~\includegraphics[width=0.225\textwidth]{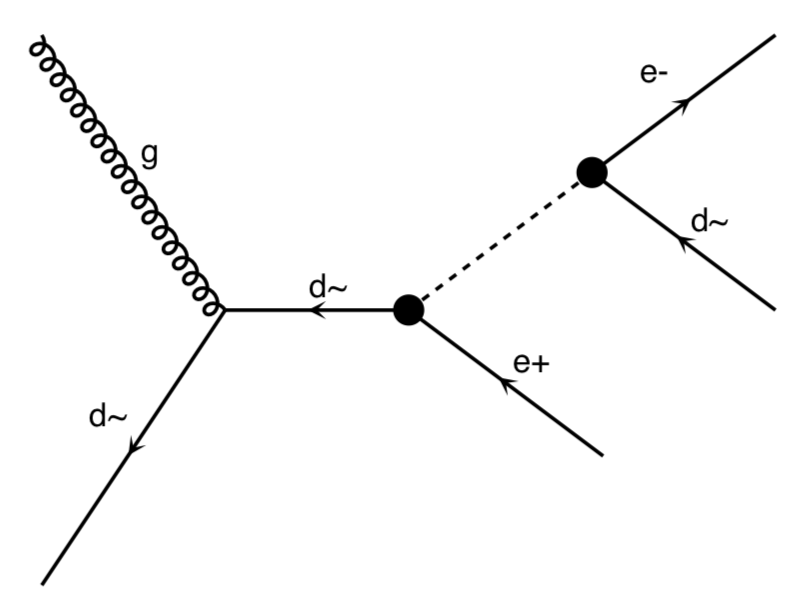}
\caption{Sample of diagrams generated by \textsc{MadGraph5\_aMC@NLO}
  in the NLO evaluation of the Drell-Yan production. The leptoquark is
  represented by the dashed line.}
  \label{fig:diag}
\end{figure}

In order to confront the leptoquark predictions with the available
data, first we simulate the processes~(\ref{eq:dy}) and
(\ref{eq:monoj}) using
\textsc{MadGraph5\_aMC@NLO}~\cite{Alwall:2014hca} with the UFO files
for our effective lagrangian generated with
\textsc{FeynRules}~\cite{Christensen:2008py, Alloul:2013bka}.
 The counter-terms needed for the NLO evaluation of the Drell-Yan
  process were evaluated using the package
  \textsc{NLOCT}~\cite{Degrande:2014vpa}.  Moreover, in NLO
  calculations, \textsc{MadGraph5\_aMC@NLO} automatically includes not
  only the SM one-loop and real emission contributions, but also the
  $t$-channel leptoquark exchange at tree level (see first diagram of
  Fig.~\ref{fig:diag}), as well as the virtual corrections involving
  leptoquarks (as in the second and third diagrams in
  Fig.~\ref{fig:diag}) and its contribution to the real emission (last
  diagram of Fig.~\ref{fig:diag}).  For further details,
  see~\cite{AlexMSc, Dorsner:2018ynv}.\smallskip


The 8 TeV Drell-Yan simulation was carried out in next-to-leading
order QCD at the parton level while the 13 TeV dilepton analyses were done
in leading order using \textsc{PYTHIA}~\cite{Sjostrand:2006za} to
perform the parton shower, while the fast detector simulation was
carried out with \textsc{Delphes}~\cite{deFavereau:2013fsa}.
In the jet plus missing transverse momentum analysis we merged the
Monte Carlo simulations containing 0, 1 and 2 jets and performed the 
jet matching.
In order to account for higher order corrections and additional
detector effects we normalized bin by bin our dilepton simulations to
the corresponding ones performed by the experimental collaborations.
Then we apply these correction factors to our simulated distributions
taking into account the leptoquark contributions. \smallskip

In order to perform the statistical analysis we defined a $\chi^2$
function for the dilepton analysis given by\footnote{In the
    Drell-Yan analysis, we estimated the effect of the systematic
    uncertainties by means of a simplified treatment in terms of two
    pulls~\cite{Fogli:2002pt, GonzalezGarcia:2007ib}. }
\begin{equation}
\chi_{\ell\ell}^2(m_R, \lambda) = \sum_i \frac{(N^i_{\rm sig}(m_R,
  \lambda)+N^i_{\rm bkg}-N_{\rm obs}^i)^2}{N^i_{\rm obs}+\sigma_{\rm
    bkg}^{i2}} \;\;,
\label{eq:chi2dilepton}
\end{equation}
where $N_{\rm sig}$ are the signal events obtained for the different
models using the corrections factors and $\sigma^i_{\rm bkg}$ are the
uncertainty on the background obtained by the experiments. Here $m_R$
($\lambda$) stands for the leptoquark mass (Yukawa coupling).
On the other hand, for the monojet searches we used a gaussian
likelihood containing the correlation between the different bins
obtained in Ref.~\cite{CMS:2017tbk}
\begin{equation}
\chi_{\rm monojet}^2(m_R, \lambda) =\sum_{i,j} (N^i_{\rm sig}(m_R,
\lambda)+N^i_{\rm bkg}-N^i_{\rm obs}) (\sigma^2)_{ij}^{-1}
(N^j_{\rm sig}(m_R,\lambda)+N^j_ {\rm bkg}-N^j_{\rm obs}) \;\;,
\label{eq:chi2monojet}
\end{equation}
where $\sigma_i^2= \sigma_i \rho_{ij}\sigma_j$ and $\rho$ is the
correlation matrix. \smallskip
Finally, the combined results are obtained with the following $\chi^2$:
\begin{equation}
\chi_{\rm combined}^2(m_R, \lambda) =\sum_{i} \chi^2_i(m_R, \lambda) \;\;,
\label{eq:chi2combined}
\end{equation}
where $i$ runs over the different searches (Drell-Yan, dilepton and
monojet).\smallskip

\section{Results}
\label{sec:results}

\begin{figure}[t!]
\centering
\includegraphics[width=0.45\textwidth]{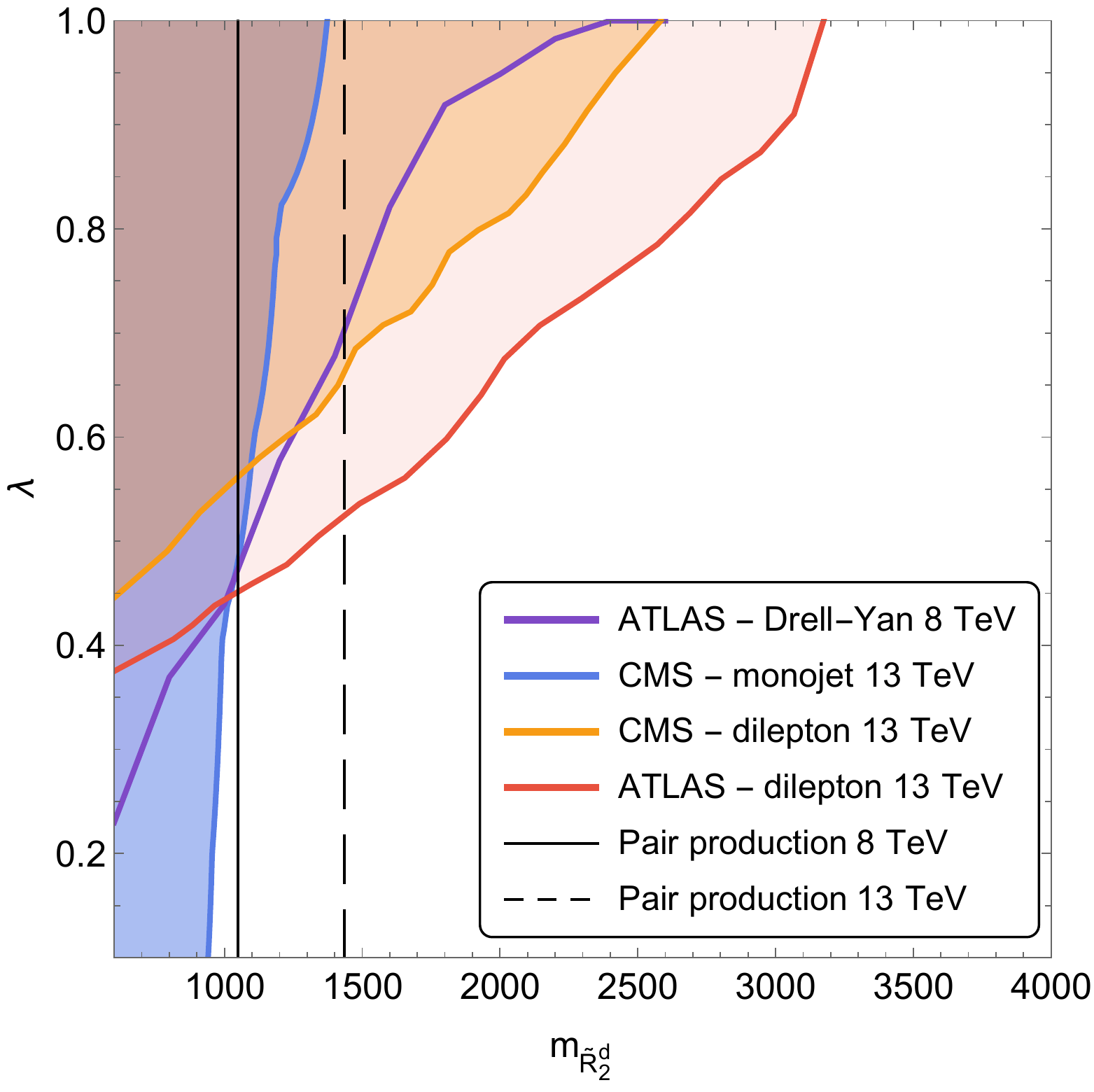}
~\includegraphics[width=0.45\textwidth]{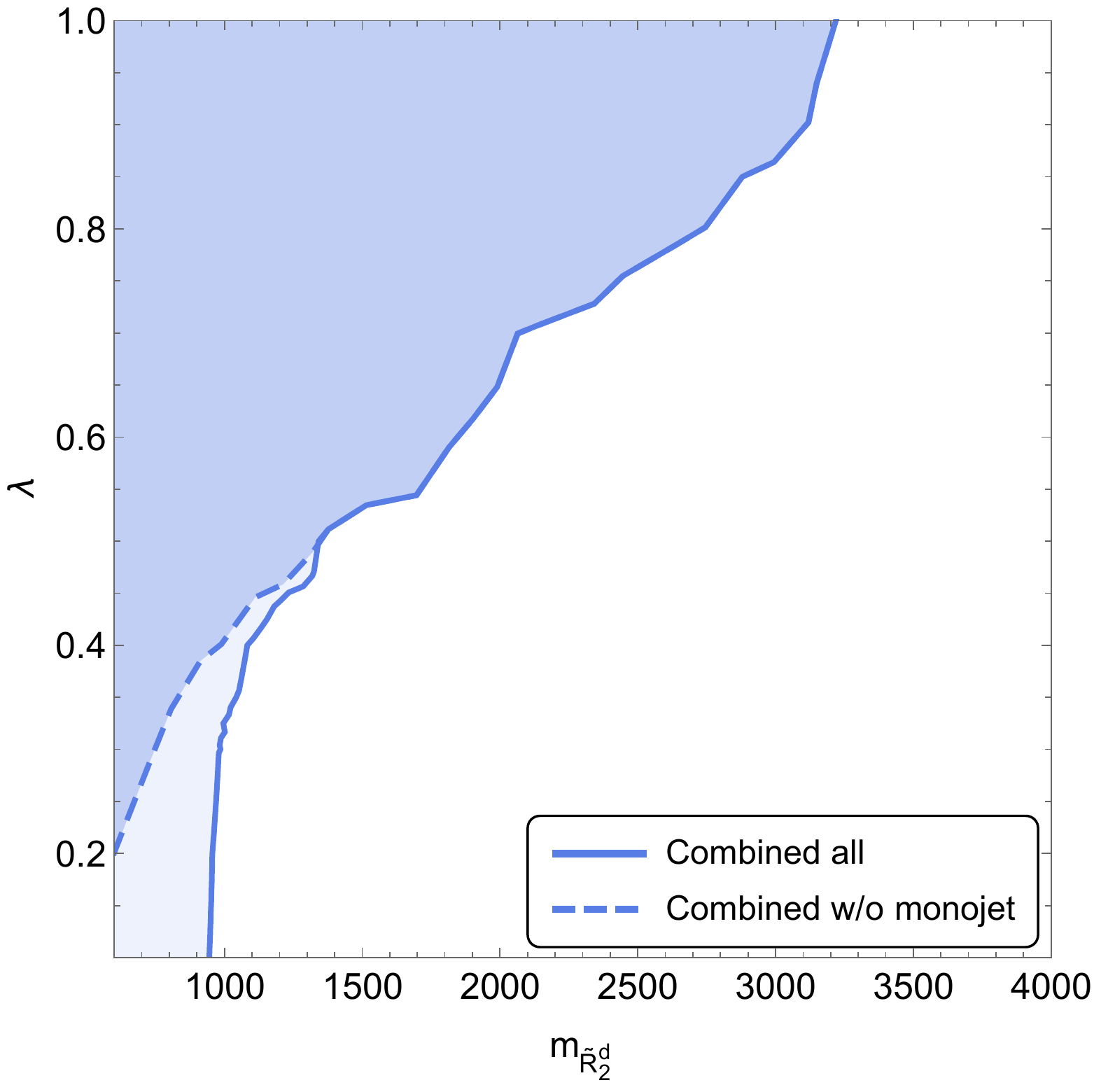}
\caption{ Excluded regions at 95\% CL for the $\tilde{R}_2^d$
  scenario. \emph{Left:} limits from the ATLAS Drell-Yan search at 8
  TeV (purple)~\cite{Aad:2016zzw}, the CMS monojet search at 13 TeV
  (blue)~\cite{CMS:2017tbk}, the CMS and ATLAS dilepton searches at 13
  TeV (yellow and red, respectively)~\cite{Aaboud:2017buh,
    Sirunyan:2018exx}, and the pair production experimental
  searches at 8 and 13 TeV (black solid and black dashed,
  respectively)~\cite{Aad:2015caa,
    Khachatryan:2015vaa,Sirunyan:2018btu,
    Sirunyan:2018ryt}. \emph{Right:} combined limits including all
  searches (solid blue) and excluding only the monojet one (dashed
  blue).}
  \label{fig:r2d}
\end{figure}

We start our analyses considering the first-generation scenarios.  In
Figure~\ref{fig:r2d} we present the results of our analyses for the
$\tilde{R}^d_2$ case. The left panel of this figure depicts the
excluded regions at 95\% CL for each of the four data sets described
in the previous section. The shaded purple region on the left of the
curve is excluded by the Drell-Yan analysis of the ATLAS 8 TeV
search. The blue area is excluded by the CMS monojet analysis at 13
TeV, while the CMS and ATLAS dilepton analysis of 13 TeV data exclude
the yellow and red regions, respectively. Finally, for the sake
  of comparison, the solid and dashed black lines show the exclusion
from the pair production experimental searches at 8 and 13 TeV,
respectively.  As we can see, the most stringent limits from masses
above 1 TeV originates from the ATLAS data on the search of new
resonances decaying into $e^+e^-$ pair~\cite{Aaboud:2017buh}.  On the
other hand, in the low mass region ($\simeq 1$ TeV) the monojet data
leads to stronger limits on the leptoquark
$\tilde{R}_2$ than the dilepton analysis. \smallskip

When we combine the three datasets on dileptons, as shown by the dashed 
blue curve in the right panel of Fig.~\ref{fig:r2d}, we see that the bounds are 
basically due to the ATLAS 13 TeV data on dileptons, except for masses 
smaller than 700 GeV where the stronger limits come from the 8 TeV Drell-Yan
analysis. Furthermore, the inclusion of the monojet data in the
combined $\chi^2$ strengthen the limits for low leptoquark masses, as shown by the 
solid blue curve. It is important to keep in mind that the combination of dilepton 
and monojet constraints is only possible because we assumed that the leptoquarks
belonging to the multiplet $\tilde{R}_2$ are degenerate.\smallskip

\begin{figure}[t!]
\centering
\includegraphics[width=0.45\textwidth]{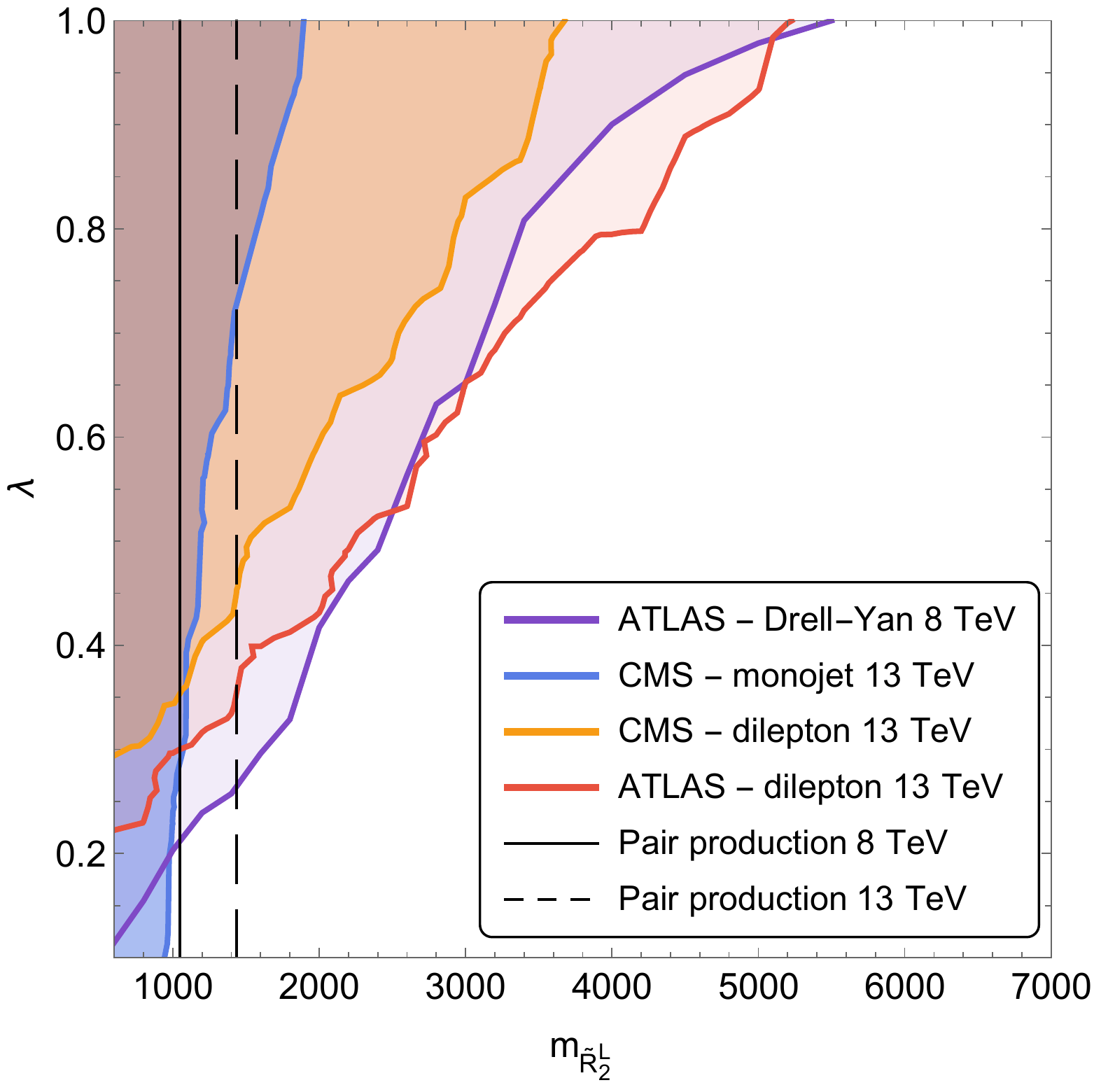}
~\includegraphics[width=0.45\textwidth]{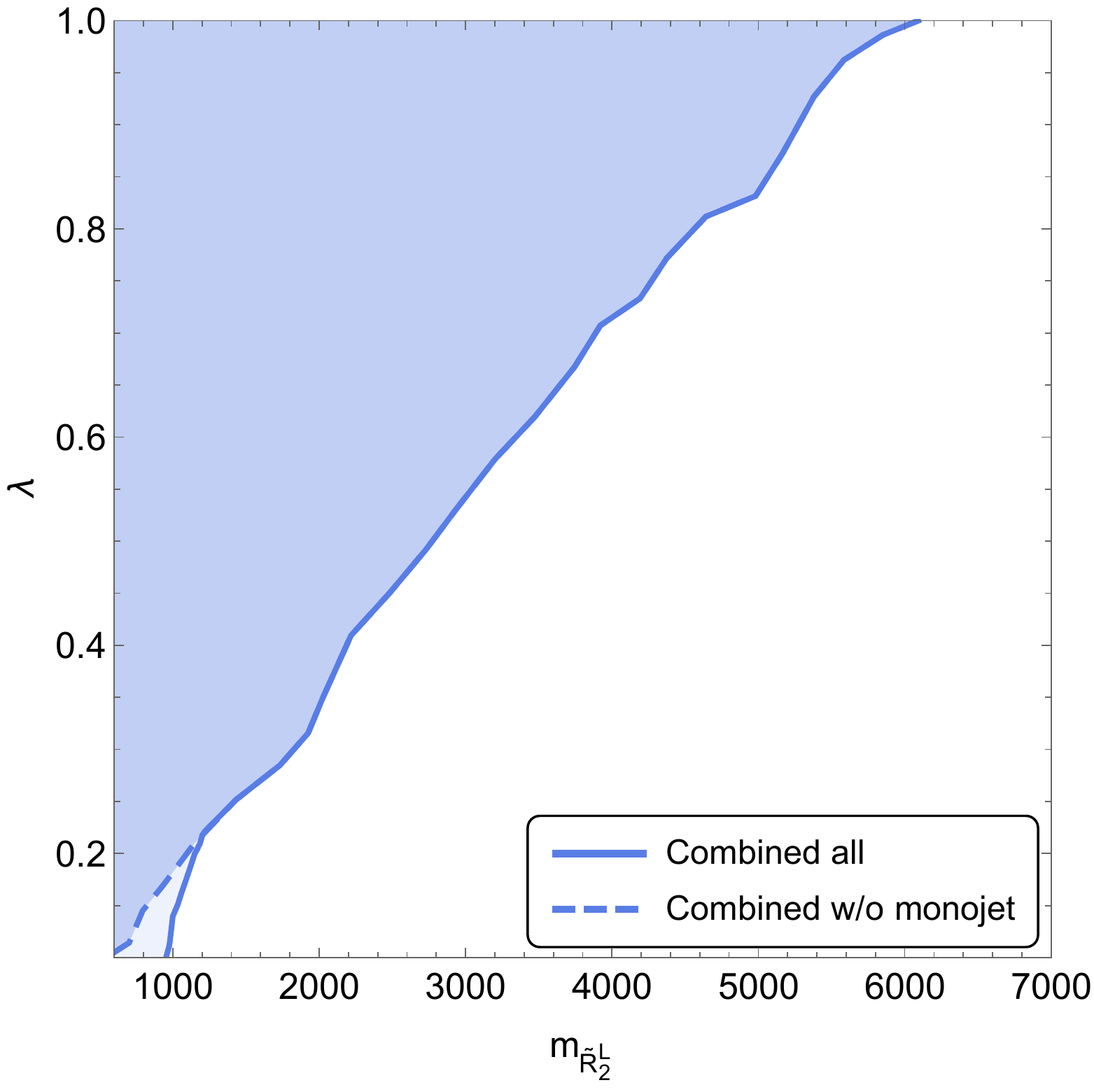}
\caption{ Excluded regions at 95\% CL for the $R_2^L$ scenario.
  \emph{Left:} limits from the ATLAS Drell-Yan search at 8 TeV
  (purple)~\cite{Aad:2016zzw}, the CMS monojet search at 13 TeV
  (blue)~\cite{CMS:2017tbk}, the CMS and ATLAS dilepton searches at 13
  TeV (yellow and red, respectively)~\cite{Aaboud:2017buh,
    Sirunyan:2018exx}, and the pair production experimental
  searches at 8 and 13 TeV (black solid and black dashed,
  respectively)~\cite{Aad:2015caa,
    Khachatryan:2015vaa,Sirunyan:2018btu,
    Sirunyan:2018ryt}. \emph{Right:} combined limits of the Drell-Yan
  and dilepton searches.). \emph{Right:} combined limits including all
  searches (solid blue) and excluding only the monojet one (dashed
  blue).  }
  \label{fig:r2l}
\end{figure}

We expect stronger constraints on the $R_2^L$ scenario than on the
$\tilde{R}_2^d$ one since the former couples to up quarks while the
latter to down quarks. This indeed happens as shown in
Figure~\ref{fig:r2l}. In the left panel the purple, blue, yellow and
red shaded regions correspond respectively to 95$\%$ CL limits from 8
TeV ATLAS Drell-Yan, 13 TeV CMS monojet, 13 TeV CMS dilepton and 13
TeV ATLAS dilepton searches. Moreover, the black solid and dashed
lines denotes the 95$\%$ CL experimental constraints from pair
production at 8 and 13 TeV. \smallskip

We learn from the left panel of this figure that for masses below 1
TeV the monojet data give rise to the tighter bounds than the
  dilepton study. On the other hand, for larger masses the ATLAS
dilepton data at 13 TeV and the ATLAS 8 TeV Drell-Yan data lead to
similar constraints. Despite the lower integrated luminosity and
energy the 8 TeV Drell-Yan limits are similar to the 13 TeV ones since
the Drell-Yan cuts and presentation of the data are more suitable for
the present study. \smallskip

The right panel of Fig.~\ref{fig:r2l} contains the limits obtained combining 
the dilepton datasets only (dashed blue curve) as well as limits 
including the monojet one (solid blue line). In this
scenario, the combined dilepton bounds are tighter then the separate
limits. Moreover, the inclusion of the monojet data give rise to
stronger limits for masses below 1 TeV. \smallskip

\begin{figure}[t!]
\centering
\includegraphics[width=0.45\textwidth]{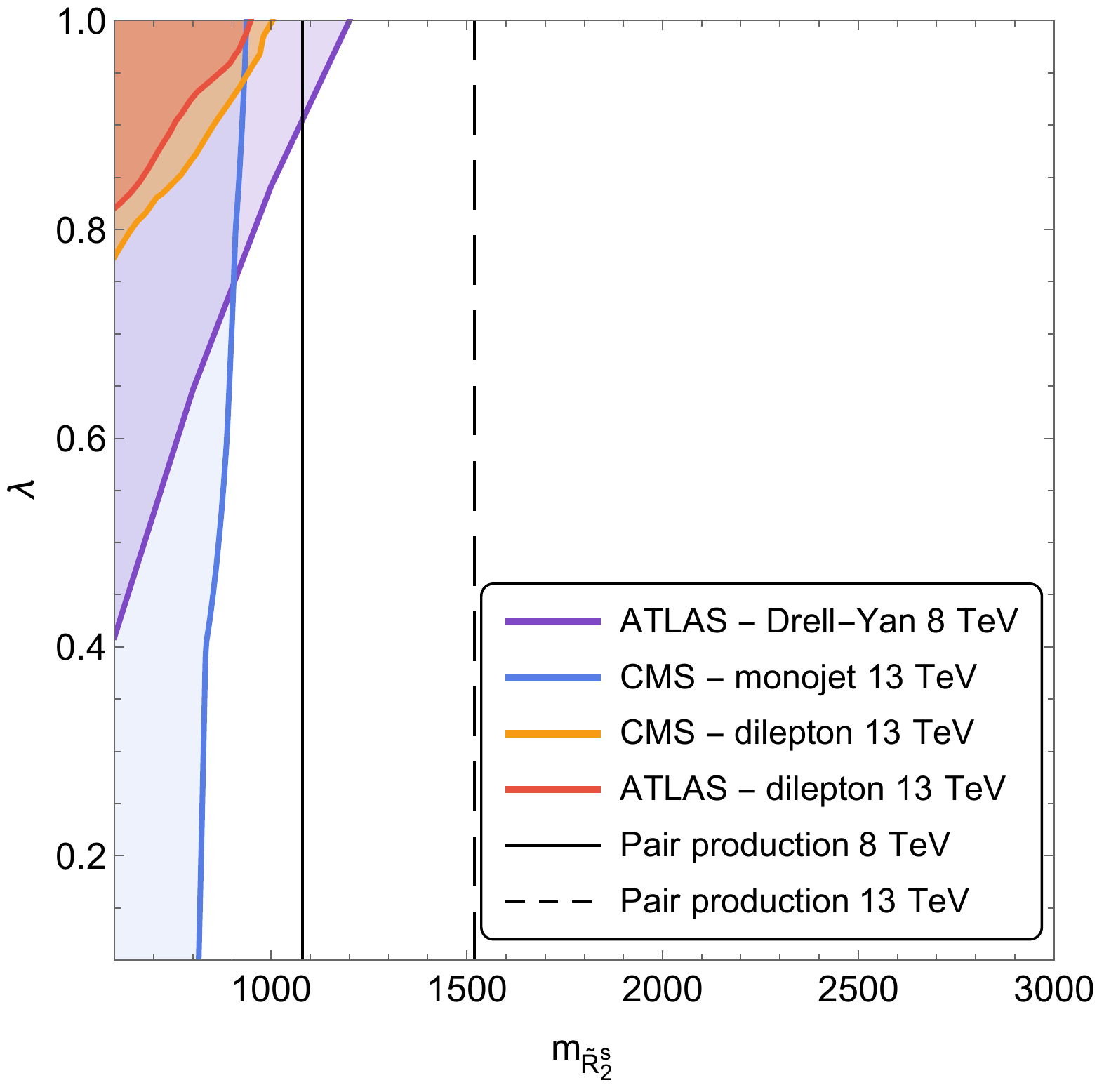}
~\includegraphics[width=0.45\textwidth]{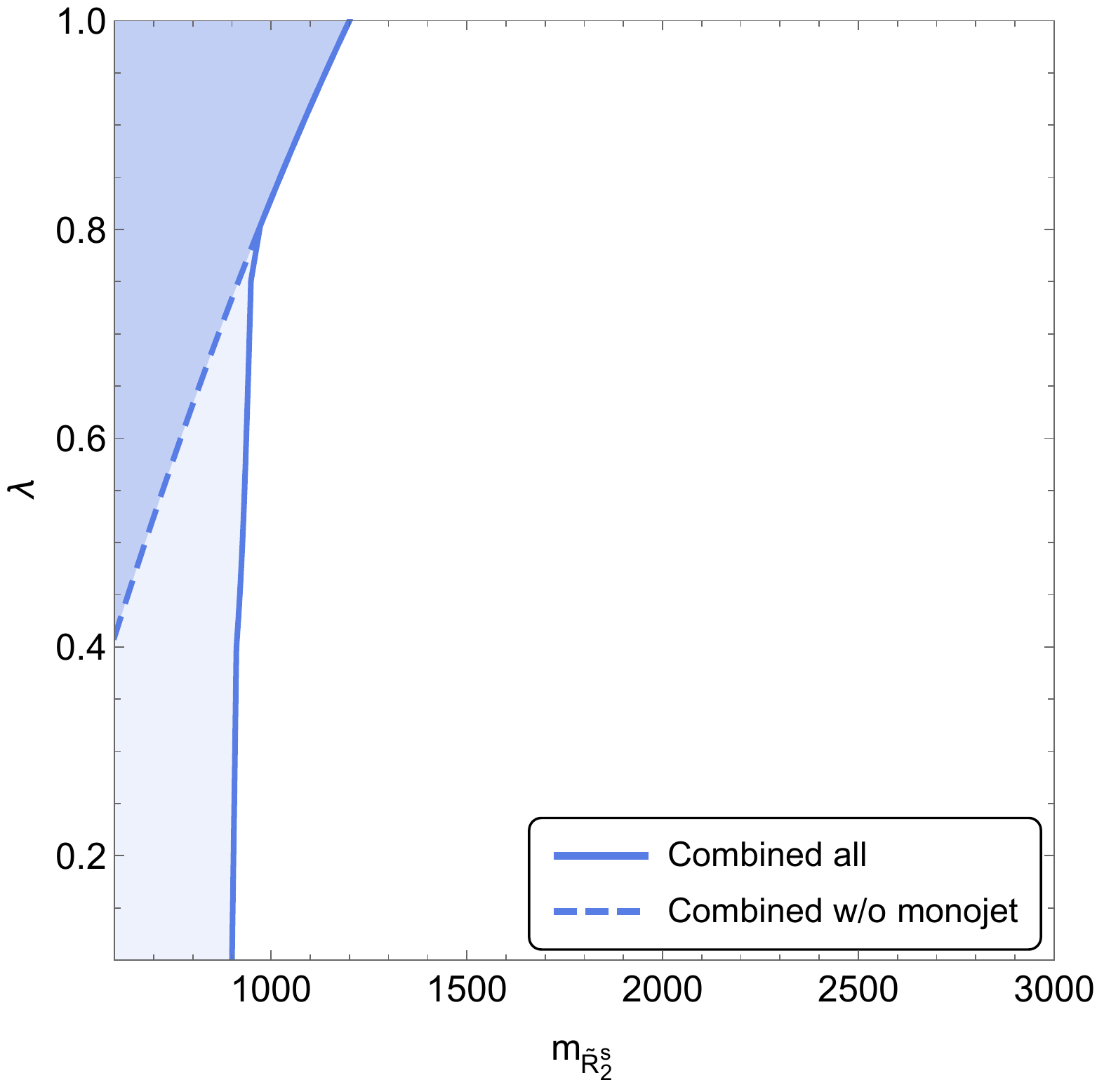}
\caption{ Excluded regions at 95\% CL for the $\tilde{R}_2^s$
  scenario. \emph{Left:} limits from the ATLAS Drell-Yan search at 8
  TeV (purple)~\cite{Aad:2016zzw}, the CMS monojet search at 13 TeV
  (blue)~\cite{CMS:2017tbk}, the CMS and ATLAS dilepton searches at 13
  TeV (yellow and red, respectively)~\cite{Aaboud:2017buh,
    Sirunyan:2018exx}, and the pair production experimental
  searches at 8 and 13 TeV (black solid and black dashed,
  respectively)~\cite{Aad:2015caa,
    Khachatryan:2015vaa,Sirunyan:2018btu,
    Sirunyan:2018ryt}. \emph{Right:} combined limits including all
  searches (solid blue) and excluding only the monojet one (dashed
  blue).}
  \label{fig:r2s}
\end{figure}

The scenario $\tilde{R}^s_2$ provides the strongest limits on second
generation leptoquarks due to its coupling to strange quarks. We
display in Figure~\ref{fig:r2s} the results for this scenario, with the 
same color coding of the previous Figures. The
most stringent dilepton limits come from the 8 TeV Drell-Yan analysis
with the ATLAS and CMS 13 TeV data leading to similar weaker constraints; see
the left panel of this figure. Moreover, the impact of the monojet is
similar to the ones of the other cases: the monojet data
leads to the best bounds for small to moderate Yukawa couplings, due to the small 
dilepton cross section at small couplings. This fact can be clearly seen on the left 
panel of this figures that depicts the results combining only the dilepton 
searches (blue dashed curve) or including the monojet dataset (solid blue line). 
The left panel shows also that, for this scenario, the pair production searches still 
give the strongest constraints.  
\smallskip

\section{Discussion}
\label{sec:discussion}


The indirect leptoquark signal in the Drell-Yan process is
complementary to the pair production searches since it is able to
probe higher leptoquark masses. On the other hand, the indirect
analysis requires the introduction of the unknown leptoquark Yukawa
coupling.
To illustrate this point let us focus on the 8 TeV data and
first-generation leptoquarks. From the left panel of
Fig.~\ref{fig:r2d} we see that the analysis of indirect effects on the
Drell-Yan production expands the excluded region by the pair production
search for Yukawa couplings $\lambda \gtrsim 0.5$. In fact, the 8 TeV indirect
limits are still more stringent than the direct searches at 13 TeV for Yukawa 
couplings $\lambda \gtrsim 0.7$.
Moreover, when we consider all available 13 TeV data, the indirect
limits are stronger than the pair production ones for Yukawa couplings
$\lambda\gtrsim 0.5$ and extend the exclusion region to leptoquark masses 
of the order of 3.5 TeV if the leptoquark Yukawa coupling is $\lambda\sim1$. \smallskip

The importance of the indirect effects is even more dramatic when we
consider the $R_2^L$ scenario where the scalar leptoquark couples to
up quarks. Indeed, Fig.~\ref{fig:r2l} shows that the 8 TeV Drell-Yan
limits are more stringent than the presently available 13 TeV direct
search limits for Yukawa couplings $\lambda\gtrsim 0.25$.
In addition, the combined 8 and 13 TeV dilepton data allow the
exclusion of leptoquarks with mass smaller than 7 TeV provided the
Yukawa coupling is $\lambda\sim1$.
In the case of second generation leptoquarks, scenario
$\tilde{R}_2^s$, the impact of the leptoquarks in the dilepton
invariant mass spectrum is much smaller due to its coupling to strange
quarks. Therefore, the indirect limits turn out to be milder than the
pair production ones that proceed through strong interactions; see
Fig.~\ref{fig:r2s}. \smallskip

At this point it is interesting to compare our findings with previous
works that analyze indirect effects of leptoquarks at the LHC. For
very large leptoquark masses their exchange in the $t$-channel can be
encoded by four-fermion dimension-six operator~\cite{deBlas:2013qqa,Bessaa:2014jya,Greljo:2017vvb}.
This method has two shortcomings since it does not take into account
properly the leptoquark interference with the SM contribution, as well
as, the leptoquark propagator~\cite{Bessaa:2014jya}. In any case, the
available limits using this procedure are much weaker than then ones
presented here. These drawbacks can be mitigated with the addition of
form factors to the four-fermion contact
interactions~\cite{Davidson:2014lsa} leading to similar results to the
ones here obtained for the Run 1 data.  However, for a specific model
it is better to fit it directly to data avoiding any ambiguity and
also allowing the evaluation of higher order corrections.  \smallskip


Our monojet analyses lead to limits that complement the dilepton
studies at low leptoquark masses (smaller than 1 TeV). For large
Yukawa coupling, {\em e.g.} $\lambda\sim1$, the monojet analysis exclude
leptoquarks with masses up to 2 TeV (950 GeV) for first- (second-)
generation leptoquarks. 
Notwithstanding, for this region of Yukawa couplings the indirect
dilepton process leads to much stronger constraints. 
Furthermore, our results indicate that the monojet channel and the
single production of leptoquarks decaying into a charged lepton and a
jet~\cite{Khachatryan:2015qda} have similar capability to search for
these particles. 
On the other hand, for second generation leptoquarks the monojet
search is not competitive with respect to the dilepton
analysis. \smallskip

Confronting the monojet bounds on first-generation leptoquarks with
ones originating from the ongoing studies of the $\ell^+ \ell^- jj$
and $\ell^\pm \nu jj$ topologies~\cite{Sirunyan:2018btu}, we learn
that the monojet search leads to tighter
bounds than these pair production searches only for Yukawa couplings
$\lambda\gtrsim 0.7$ in the $\tilde{R}_2^L$ scenario. 
Furthermore, our results show that the monojet channel extends
considerably the exclusion limits originating from leptoquark pair
production followed by their decay into a jet and a neutrino, {\em
  i.e.} the $jj E^{\rm miss}_T$ channel~\cite{Sirunyan:2018kzh}.
\smallskip


We can also abandon the hypothesis of degeneracy of leptoquarks belonging
to the same multiplet if we consider separately the dilepton and
monojet constraints. In this case, we have to examine the couplings of
each component of the leptoquark multiplets:
\begin{enumerate}
\item the dilepton bounds on $\tilde{R}_2^d$ applies to all leptoquarks that couples
to electrons and down quarks, that is, $\tilde{S}_1$, the components 1
and 2 of $\vec{S}_3$, the down component of $R_2$ with right-handed
couplings, and the up component of $\tilde{R}_2^d$ (of course);
\item the dilepton constraints on $R_2$ are the same for
$S_1$, the third component  of $\vec{S}_3$ and the up (down) component of
$R_2$ with left-handed (right-handed) coupling;
\item the monojet analyses labeled $\tilde{R}_2^d$ is the same for $S_1$,
the third component of $\vec{S}_3$ and the down component of
$\tilde{R}_2^d$. Our $R_2^L$ monojet study is valid for the components 1
and 2 of $\vec{S}_3$ and the down component of $R_2$ with left-handed
couplings.
\end{enumerate}\smallskip

Finally, let us comment on models containing leptoquarks that
  address the anomalies observed in lepton flavour universality. In
  general, see for instance Ref.~\cite{Becirevic:2016oho}, the
  proposed ${\cal O}$(TeV) leptoquarks mediate transitions between the
  second and third families due to strong $e-\mu$ conversion
  constraints. Our results for the $\tilde{R}^s_2$ scenario indicate
  that indirect effect on the $\mu^+ \mu^-$ production will not
  contribute to strengthen the direct pair searches for scalar
  leptoquarks.  Some models also present vector leptoquarks. However,
  even the direct search limits on them depend upon two additional
  parameters needed to describe their $SU(3)_C$ gauge interactions,
  therefore rendering the analyses much more
  involved~\cite{Blumlein:1996qp, Belyaev:1998ki}. High-$p_T$
  observables involving the processes $pp\to\tau\tau$ and $pp\to
  \tau\mu$ are more sensitive to leptoquarks mediating interactions
  with the third generation which are also relevant to the observed
  $B$-anomalies~\cite{Angelescu:2018tyl,Baker:2019sli, Cornella:2019hct}. The relevant
  monojet channel, by its turn, might involve bottom jets and should
  benefit more from a dedicated analysis for $b$-tagged
  jets~\cite{Lin:2013sca}.
\smallskip

\section{Conclusions}
\label{sec:conclusions}

In this work, we have considered collider constraints on first- and
second-generation leptoquarks. In particular, we studied the
$\tilde{R}_2$ scalar leptoquark model with couplings only to the
right-handed down (strange) quark and the first- (second-) generation
lepton doublet, and the $R_2$ scalar leptoquark model with couplings
only to the right-handed up quark and the first generation left-handed
lepton doublet. Stringent experimental bounds on leptoquarks come from
their pair production followed by their prompt decays. This production
mechanism is independent of the leptoquarks Yukawa coupling. For large
leptoquark Yukawa couplings, on the other hand, single
production becomes important. This is especially relevant for couplings
with the first generation quarks because of their large parton
distribution functions. \smallskip

Our results show the importance of the search for leptoquark effects
in the Drell-Yan process since it allows to extend considerably the
reach in leptoquark mass for moderate to large leptoquark Yukawa
couplings. For example, the scalar $\tilde{R}_2$ ($R_2$) leptoquark
model with couplings to the first-generation quark is excluded up to
$\sim 3.2$ ($\sim5.6$) TeV for $\lambda\sim1$, overcoming the pair
production limits $\lesssim 1.5$ TeV.  In addition, we showed that the
monojet channel is a viable alternative to further probe
leptoquarks. Moreover, should a signal arise at the LHC, these kind of
processes would give more information on the leptoquark properties,
given their dependence on the leptoquark mass and Yukawa
coupling.\smallskip

Finally, we discussed how to interpret the results obtained in this work if we abandon 
the hypothesis of degeneracy of the leptoquarks belonging to the same $SU(2)$ 
multiplet and the possible connection to constraints on the singlet ($S_1$) or 
each component of the triplet ($S_3$) scalar leptoquark models. \smallskip

\acknowledgments 

This work is supported in part by Conselho Nacional de Desenvolvimento
Cient\'{\i}fico e Tecnol\'ogico (CNPq) and by Funda\c{c}\~ao de Amparo
\`a Pesquisa do Estado de S\~ao Paulo (FAPESP) grant 2012/10095-7.
A. Alves thanks Conselho Nacional de Desenvolvimento Cient\'{\i}fico
(CNPq) for its financial support, grant 307265/2017-0.  R.R. Moreira
was supported by FAPESP process no. 2013/26511-1.  GGdC has been
supported by the FAPESP process no. 2016/17041-0 and by the National
Science Centre, Poland, under research grant no. 2017/26/D/ST2/00225.
GGdC would like to thank FAPESP grant 2016/01343-7 for funding his
visit to ICTP-SAIFR during November 2018 where part of this work was
done.


\bibliography{referencesv4}
\end{document}